\documentclass{ws-p8-50x6-00}
\usepackage{psfig}

\begin{document}

\title{Wide Field Imaging at 250 GHz}

\author{C.L. Carilli}

\address{NRAO}

\author{F. Bertoldi,  A. Bertarini, K.M. Menten, E. Kreysa, R. Zylka}

\address{MPIfR}  

\author{F. Owen, M. Yun}

\address{NRAO}  


\maketitle

\abstracts{
We summarize results from sensitive, wide-field imaging 
using the Max-Planck Bolometer Array at the IRAM 30m
telescope, including source counts, clustering, and redshift
distribution. 
}

%\section{}

The Max-Planck mm Bolometer Array (MAMBO) is a 37 element, 250 GHz bolometer
array operating at 300 mK.  We have observed extensively with 
MAMBO at the IRAM 30m telescope over the last two winters,
including pointed observations of selected samples of high
redshift objects to typical rms sensitivities of 0.3 mJy, and 
wide field raster imaging to sensitivities of about 0.5 mJy over 
areas $\sim 200$ arcmin$^{2}$. The pointed observations will be presented
elsewhere. In this short contribution we summarize
preliminary results from the wide field imaging programs. 

The three fields observed thus far are the NTT Deep field, the Lockman 
Hole, and the $z = 0.25$ cluster Abell 2125.
In parallel with the wide field imaging at 250 GHz, we have observed,
or will be observing, these fields with the VLA at 1.4 GHz 
to rms $= 7~ \mu$Jy, and we are obtaining sensitive
optical and near-IR wide field images with various optical
telescopes. 

Figure 1 shows the MAMBO image of the Abell 2125 field.  
The rms in the center of the field is about 0.5 mJy. We detect 36
sources with flux densities, S$_{250} \ge 2$ mJy in
regions with rms noise $\le 0.6$ mJy. The brightest
source is 13 mJy. This source is also detected at 1.4 GHz with
S$_{1.4} = 40$ mJy, and has a point source optical counterpart. 
We feel it likely that this is a radio-loud QSO, and that the
250 GHz emission is the high frequency extrapolation
of the AGN synchrotron spectrum. 
We have examined in detail the optical and near-IR
properties of a few of the MAMBO
sources in the Abell 2125 field\cite{bertoldi}. 
These appear to be the  same population as the SCUBA sources, in the
sense that they are typically very faint in the optical and near-IR, 
with $K \ge 20.5$.

\begin{figure}[ht]
%\vskip -1in
\hskip 0in
\psfig{figure=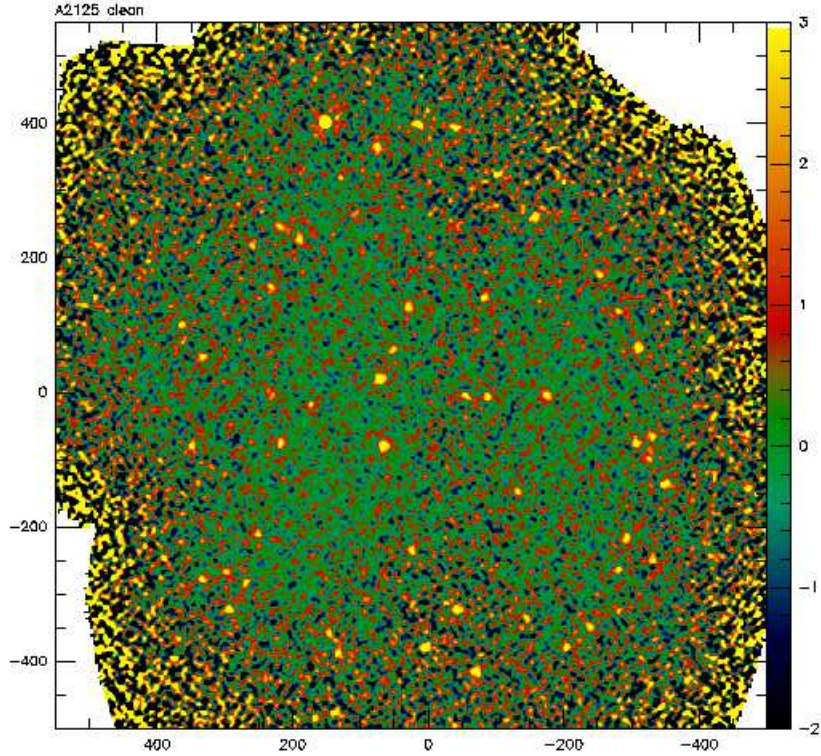,width=4.0in,angle=-90}
\caption{The MAMBO  image at 250 GHz of the Abell 2125. The 
angular scale is in arcseconds, and the greyscale
range is in mJy, and the rms = 0.5 mJy.
}
\end{figure}

Figure 2a shows the cumulative 
source counts based on two of the three MAMBO
fields, along with source counts determined from various SCUBA
surveys. We relate 250 GHz flux densities
to 350 GHz flux densities using a scaling
factor of 2.25. This factor is applicable to a typical starburst
galaxy  at $z \approx 2.5$. We have included faint source
counts in the regions within a 1$'$ radius of the cluster
center assuming a mean gravitational magnification factor of 2.5.
The MAMBO and SCUBA counts agree well at intermediate 
flux densities (S$_{350} = 2$ to 8 mJy). The very wide fields imaged by
MAMBO allow us to set the best constraints to date at high flux
densities,  and we find that there is steepening in the distribution
at S$_{350} = 10$ mJy, consistent with
an exponential cut-off in the starburst galaxy
population at about  $10^{13}$ L$_\odot$. 
All of the data can be reasonably fit by a Schechter-type
luminosity function, with a powerlaw index of --2 and an
exponential cut-off at 10 mJy.

\begin{figure}[ht]
\psfig{figure=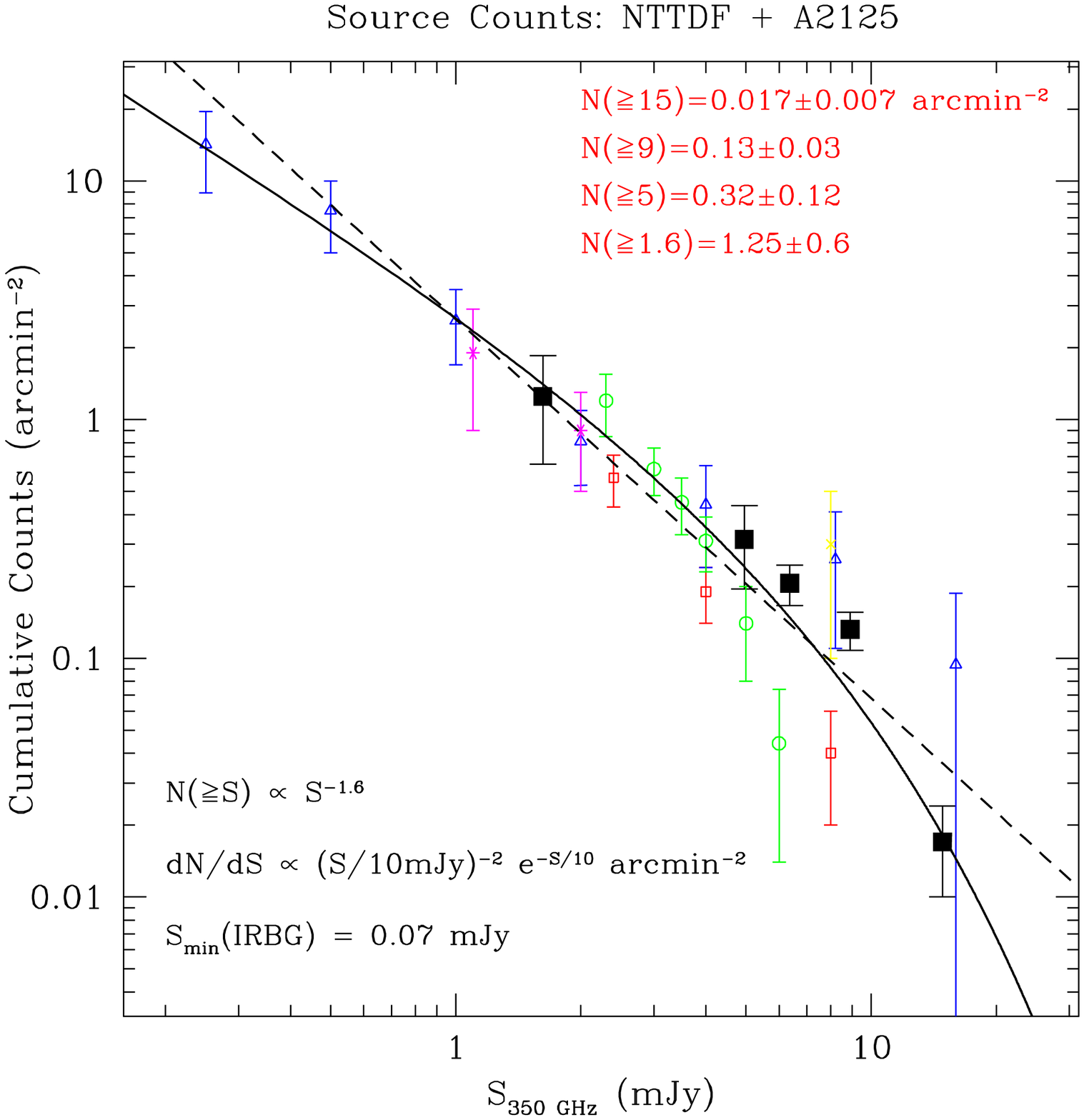,width=2.35in}
\vskip -2.35in
\hspace {2.35in}
\psfig{figure=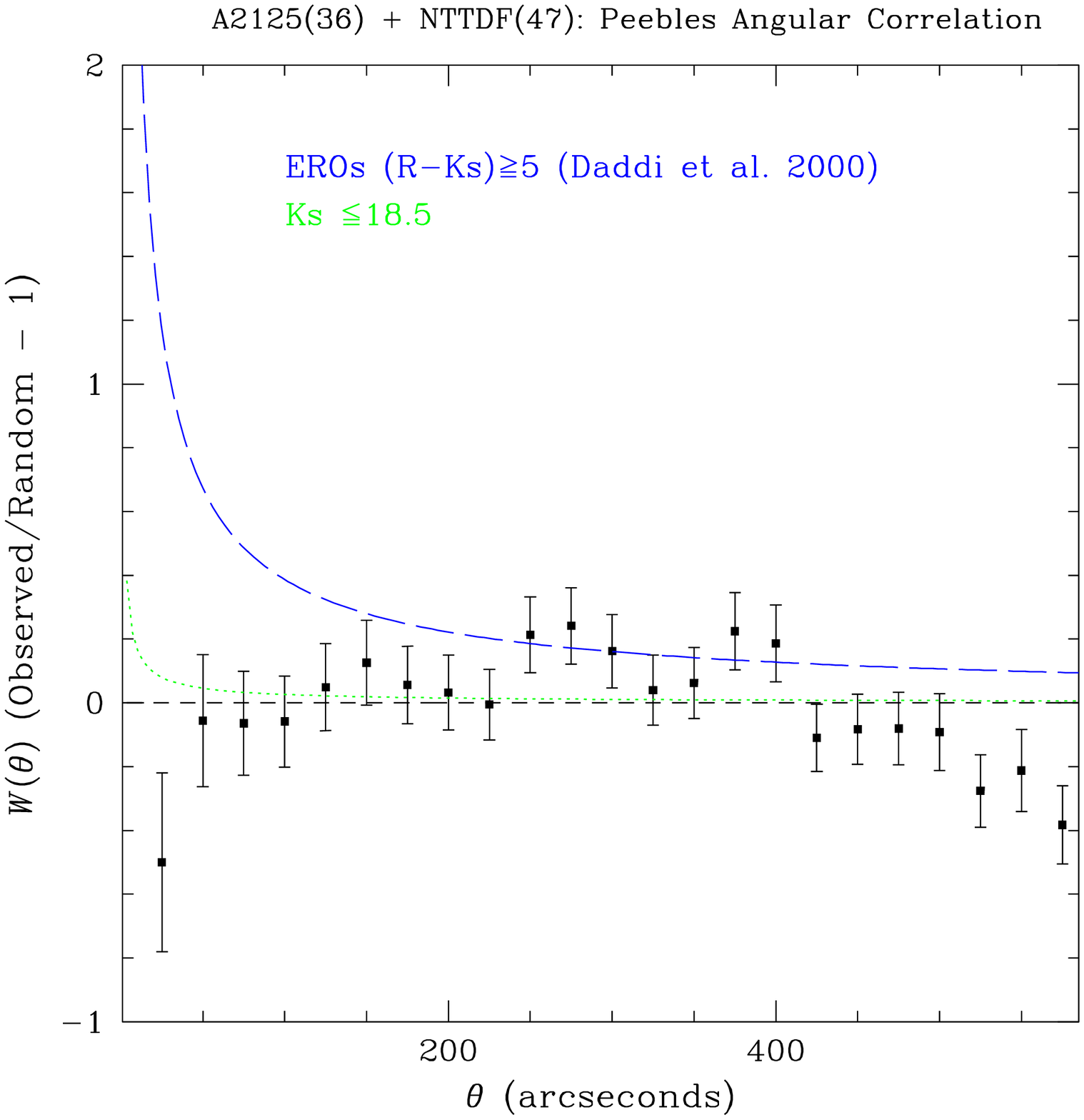,width=2.35in}
\caption{
The figure on the left shows the source counts from the MAMBO fields 
as large solid squares, plus counts from various SCUBA surveys (this
volume). The dashed curve is a powerlaw of index --1.6.  The solid
curve is an (integrated) Schechter function with parameters as given
in the text and on the plot. 
The figure on the right
shows the two point correlation function for 250 GHz sources in the 
NTT deep field and the Abell 2125 field. 
The dashed curve is for EROs and the dotted curve is for 
galaxies with $K_s \le 18.5$.
}
\end{figure}

Figure 3a shows the redshift distribution for the Abell 2125 field
sources that are detected at both 250 GHz and 1.4 GHz. We selected
this field for a redshift analysis
since it has the deepest radio image of the three MAMBO
fields, with an rms = 7 $\mu$Jy. 
The redshifts were determined using the 1.4 GHz-to-250 GHz flux density
ratio applicable to star forming galaxies\cite{cy}. We find that 18 of
the 36 sources with S$_{250} \ge 
2$ mJy are detected at S$_{1.4} \ge 22\mu$Jy. 
The median redshift is 2.5, with most of the sources between 
$2 \le z \le 4$. The high radio detection rate 
gives us confidence in the reliability of the MAMBO selected sources.
We expect only about 1 source to have been detected at random 
at 1.4 GHz within 3$''$ of any of the 36 MAMBO source positions. 
The high detection rate also implies that the sensitivities of the 
mm and cm surveys are well-matched, and that there is not a dominant
`hidden' population of sources, either at very high redshift, or 
at low redshift but with radio flux densities below those expected for
starburst galaxies based on the radio-to-far IR correlation. 
Given these results, it is likely 
that the upcoming radio imaging programs with the VLA
pushing down to an rms $\approx 3$ mJy at 1.4 GHz 
will result in a close-to-complete radio 
detection rate for fields such as this. 

Figure 3a also shows the redshift distribution for SCUBA sources at 350 GHz 
with radio detections.  This shows a lower median redshift of
1.9. A systematic offset in redshift 
between sources selected at 250 GHz relative
to 350 GHz is expected qualitatively, since the spectral energy
distributions go `over-the-top' of the IR peak at lower $z$ at 350 GHz. 
Quantitatively however, one
would expect this segregation to occur at higher redshift,
$z \approx 6$, hence we feel that the segregation in
redshift in Figure 3a is more likely due to the deeper radio
survey for Abell 2125.

\begin{figure}[ht]
\psfig{figure=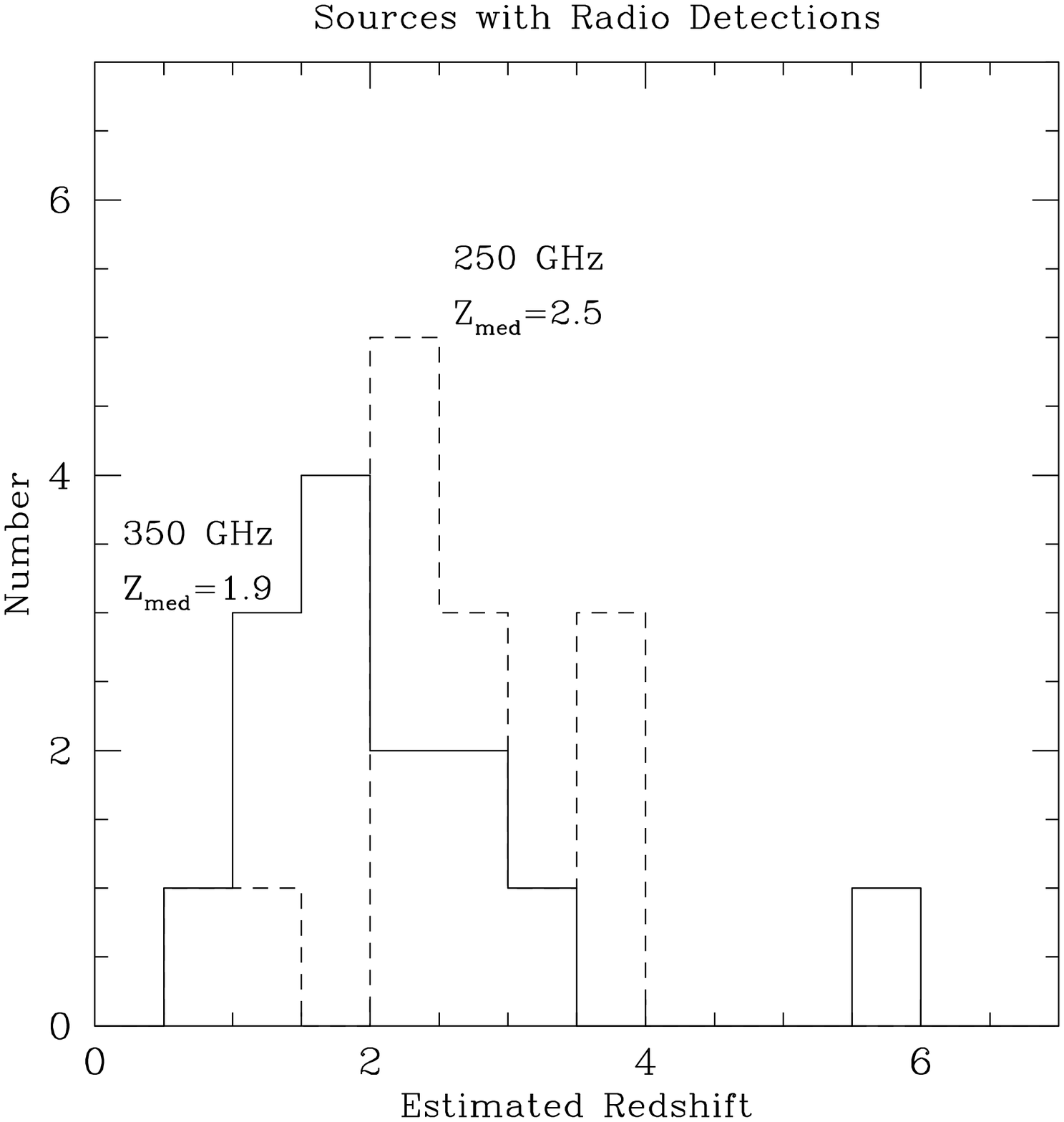,width=2.35in}
\vskip -2.35in
\hspace {2.35in}
\psfig{figure=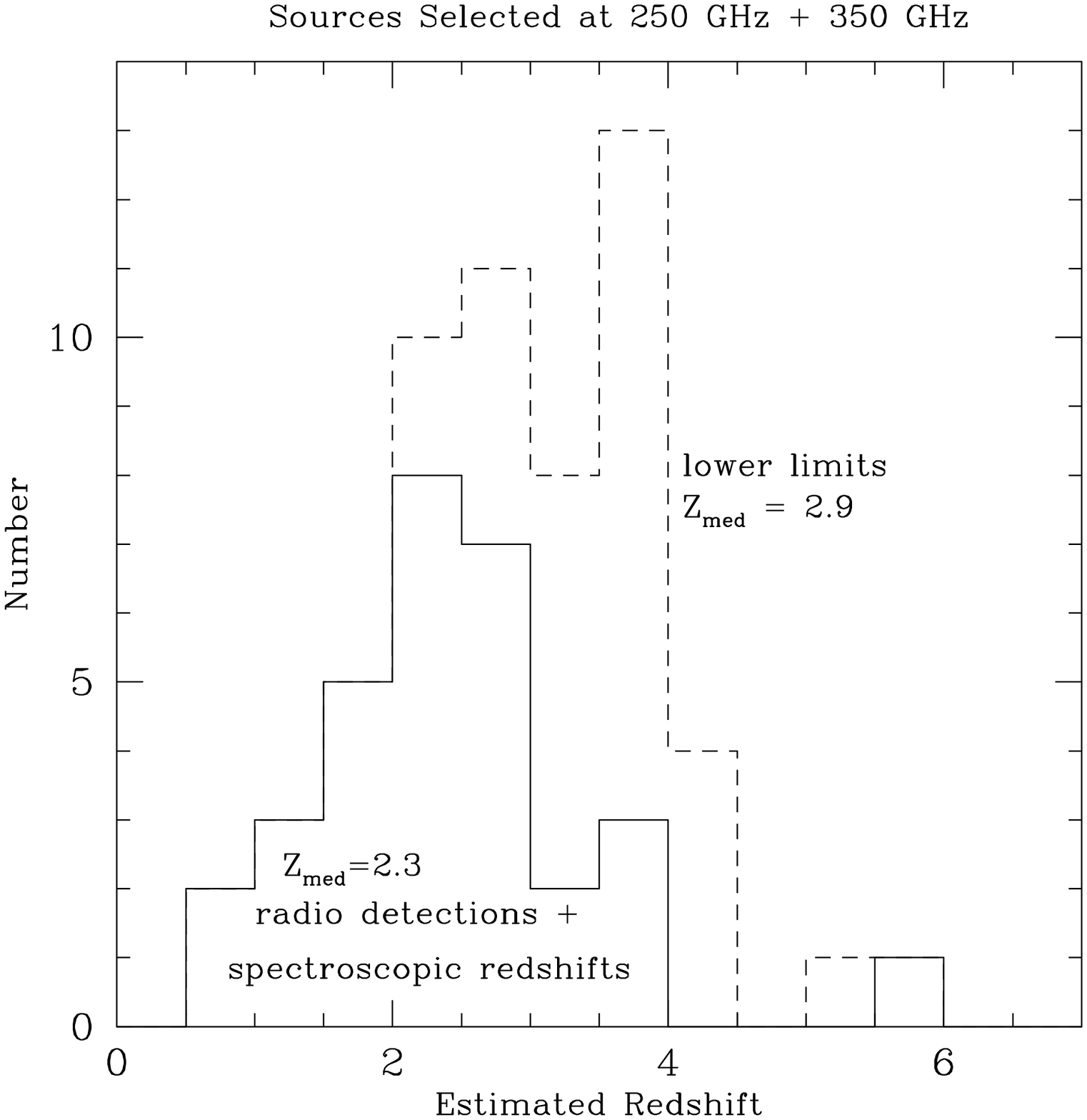,width=2.35in}
\caption{The left figure shows the redshift distribution 
for 250 GHz and 350 GHz selected sources separately, 
based on the cm-to-mm flux density ratio for sources with 
radio detections at 1.4 GHz. The right figure shows the
distribution for all the sources, including redshift lower limits
based on radio upper-limits.
}
\end{figure}

Figure 3b shows the total redshift distribution for sources 
selected at either 250 or 350 GHz, including sources with 
spectroscopic redshifts, radio detections, and radio lower limits.
The distribution is broad, but most of the sources appear to be in the
range $1.5 < z < 4$, with a median of 2.3 for sources with radio
detections and/or spectroscopic redshifts, and 2.9 if radio lower
limits are included. The lower limits leave open the possibility of a
substantial, although not majority, population of high-$z$ sources,
with $z > 3$. 

Figure 2b shows the clustering properties of the sources in 
the Abell 2125 plus the NTT deep field. Plotted is the
Peebles two-point correlation function, determined using a random
distribution of a large number of sources with the same spatial 
sampling as the MAMBO fields.  We also plot 
the clustering properties of Extremely Red Objects (EROs)\cite{Daddi}.
The results are noisy, due to the small
number of sources, but the mm-selected sources appear 
to be less clustered than the EROs. This is not
surprising, since the EROs are mostly intermediate-redshift elliptical 
galaxies with very strong clustering properties. Also,
given the very flat `redshift selection function' for mm sources
from $z = 0.5~ \rm to~ 7$ (i.e. the strong negative $K-$correction), 
any clustering will be highly diluted by the large volume sampled.

%\section*{Acknowledgments}
%The National Radio Astronomy Observatory is a facility of the Nation
%Science Foundation, operated under contract with Associated
%Univ. Inc.. 


\begin{thebibliography}{00}
%\bibitem{carilli}Carilli et al. 2000, in preparation

\bibitem{cy} Carilli, C.L. and Yun, M.S. 2000, ApJ, 530, 618

\bibitem{bertoldi} Bertoldi, F. et al. 2000, A\&A, 360, 92

\bibitem{Daddi} Daddi, E. et al. 2000, A\&A, in press

\end{thebibliography}
\end{document}